\documentclass[sigconf]{acmart}

 \renewcommand\footnotetextcopyrightpermission[1]{} 
\usepackage{booktabs} 

\usepackage{graphicx}
\usepackage{subfigure}
\usepackage{listings}
\usepackage{xcolor}

\usepackage{array}
\usepackage{url}
\newcolumntype{P}[1]{>{\raggedright\arraybackslash}p{#1}}
\usepackage{amssymb} 
\usepackage{adjustbox} 
\usepackage{colortbl} 

\definecolor{DarkGreen}{rgb}{0.0,0.8,0.0}
\newcommand{\greencheck}{\textcolor{DarkGreen}{\checkmark}}

\newcolumntype{R}[2]{%
    >{\adjustbox{angle=#1,lap=\width-(#2)}\bgroup}%
    l%
    <{\egroup}%
}
\newcommand*\rot{\multicolumn{1}{R{45}{1em}}}
\definecolor{Gray}{gray}{0.85}
\newcolumntype{g}{>{\columncolor{Gray}}c}
\newcolumntype{w}{>{\columncolor{white}}c}

\begin{document}
\title{BRISC-V: An Open-Source Architecture Design Space Exploration Toolbox \vspace{-0.0in}}

\author{Sahan Bandara, Alan Ehret, Donato Kava and Michel A. Kinsy\\
Adaptive and Secure Computing Systems (ASCS) Laboratory \\
Department of Electrical and Computer Engineering, Boston University\\
Boston, Massachusetts, 02134, USA\\
\{sahanb, ehretaj, dkava, mkinsy\}@bu.edu}

\begin{abstract}
In this work, we introduce a platform for register-transfer level (RTL)
architecture design space exploration. The platform is an open-source,
parameterized, synthesizable set of RTL modules for designing RISC-V based
single and multi-core architecture systems. The platform is designed with a
high degree of modularity. It provides highly-parameterized, composable RTL
modules for fast and accurate exploration of different RISC-V based core
complexities, multi-level caching and memory organizations, system topologies,
router architectures, and routing schemes. The platform can be used for both
RTL simulation and FPGA based emulation. The hardware modules are implemented
in synthesizable Verilog using no vendor-specific blocks. The platform includes
a RISC-V compiler toolchain to assist in developing software for the cores,
a web-based system configuration graphical user interface (GUI) and a web-based
RISC-V assembly simulator. The platform supports a myriad of RISC-V architectures,
ranging from a simple single cycle processor to a multi-core SoC with a complex
memory hierarchy and a network-on-chip. The modules are designed to support
incremental additions and modifications.
The interfaces between components are particularly designed to allow parts of the
processor such as whole cache modules, cores or individual pipeline stages,
to be modified or replaced without impacting the rest of the system.
The platform allows researchers to
quickly instantiate complete working RISC-V multi-core systems with
synthesizable RTL and make targeted modifications to fit their
needs.
The complete platform (including Verilog
source code) can be downloaded at \url{https://ascslab.org/research/briscv/explorer/explorer.html}. 
\end{abstract}


\maketitle

\section{Introduction}
Designing, building, and testing multi-core, many-core
and even single-core processor systems is a difficult and time consuming task.
Designers are faced with numerous design decisions that, when taken as a whole,
impact performance in subtle ways.
With the ever increasing size and complexity of multi-core and many-core systems,
the time and effort needed for development is quickly raising the barrier to
entry for design space exploration and research.
This growing obstacle to multi-core system design
creates a need for a flexible micro-architecture design space exploration
platform.
However, there are many challenges involved with creating such a platform.
Salient research questions related to creating such a platform include:
(1) What aspects of a multi-core system are relevant to micro-architecture
design space exploration? (2) How can a design space exploration platform
provide ease of use and rapid exploration while maintaining the speed and
accuracy of FPGA-based emulation? (3) How can the size of the design space
covered by a platform be maximized to provide support for a wide range of
systems and research?

To address these questions and challenges, we present an open-source
platform for RISC-V multi-core system micro-architecture design space exploration.
The key components of this platform are:
\begin{itemize}
  \item A modular, parameterized, synthesizable multi-core RISC-V hardware
    system written in Verilog.
  \item A RISC-V toolchain to compile a user's
    code for bare-metal execution on the hardware system.
  \item A RISC-V assembly simulator to test a user's software independently of
    the hardware system.
  \item A hardware system configuration graphical user interface to visualize
    and generate multi-core hardware systems.
\end{itemize}
The name of the platform is withheld to maintain anonymity.

Current research that would benefit from fast micro-architecture design space
exploration ranges from the development of efficient
network on-chip (NoC) \cite{3dnoc} to cache timing side channel elimination \cite{janus} to heterogeneous \cite{kumar2004single}
or adaptive architecture design \cite{angstrom}.
Indeed, there is still active research related to every subsystem in a multi-core
design.
As such, our platform supports design space exploration for:
(1) RISC-V cores, with various pipeline
depths and configurations; (2) the cache subsystem with user selectable sizes
and associativities; (3) a main memory subsystem with support for on-chip block RAM (BRAM)
or off-chip memory; and (4) a parameterized on-chip network supporting a
variety of router designs and routing algorithms.

The hardware system is written in synthesizable Verilog with no vendor specific
IP blocks. This implementation enables RTL simulation, in addition to the fast
FPGA-based emulation necessary for large, yet accurate, design space exploration. Parameterization and modularity allow for rapid changes to the design of the
system.
Individual modules (e.g. core pipeline stages or cache replacement policies) as well as
whole subsystems (e.g. core, cache or NoC designs) can be customized or replaced
independently of each other.
This allows users to make changes to their relevant systems without the need to
modify or understand implementation details of other aspects of the system.
The platform provides multiple implementations of core, cache, memory and
NoC subsystems for users to choose from.
Parameterized modules allow users to quickly change settings of the system,
such as cache size, cache associativity, or number of cores. Such parameters
enable the fine tuning of a micro-architecture after the appropriate hardware
modules have been selected.

The software tools included in the platform facilitate rapid design space
exploration by streamlining the platform work flow.
Including the necessary toolchain allows users to
develop software for their design space exploration quickly and run it on a variety of
system configurations.
The platform's RISC-V assembly simulator can be
used to create a golden model of program execution and fine tune
software before testing it on the hardware system.
These golden models can accelerate debugging efforts by providing an expected
execution flow.
The hardware system configuration GUI allows users to select core
types and features, cache sizes, and associativities, bus-based or NoC-
based interconnects, among many other parameters.  Visualizations of
the configuration promptly give users an understanding of
their system.  Users can use the GUI to generate an RTL implementation
of the system, allowing them to easily make and visualize changes,
before producing a new design.
The modularity that comes with support for interchangeable core, cache, memory,
and NoC subsystems (and their internal modules) enables a wide breadth of design
space for exploration, without any RTL modifications to the base platform. In order to maximize the explorable design space, every aspect of the platform
is open-source.
Users looking to expand the number of supported subsystems or add experimental
features can do so by modifying or extending the Verilog RTL of the base
system.

The use of RISC-V, an open ISA that is freely available, adds additional possibilities for design space exploration with
its modular nature and numerous extension specifications \cite{risc_v_spec}
\cite{risc_v_privileged_spec}. Users can add custom instruction set extensions to support experimental
architecture features or custom hardware accelerators. The openness and the option for custom extensions makes RISC-V an
excellent ISA for design space exploration. This platform uses version 2.2 of the RISC-V User-Level ISA \cite{risc_v_spec}

The complete platform (including Verilog source code) can be downloaded at \url{https://ascslab.org/research/briscv/explorer/explorer.html}.

\section{Related Work}
Other work has developed configurable processors and tools to facilitate varying degrees of design space exploration.
One such tool, Heracles~\cite{kinsy2013heracles}~\cite{kinsy2011heracles}, is
based on the MIPS ISA and runs on the Windows operating
system. Everything needed to create
a synthesizable system is included with Heracles. A GUI enables users to
specify their system and generate Verilog code for it. Two core types, each with one
or two hardware threads, can be selected for use in multi-core systems. A dummy core is included to test different on-chip network designs. Cache hierarchies can involve one or two levels of direct mapped caches. A cross compiler
allows programmers to write parallel code for a multi-core MIPS architecture.
A wide variety of NoC routing configuration options are available in the NoC
configuration of Heracles.

Heracles and our platform share similar goals, however, Heracles is more limited
in the number of core and cache configurations available. Additionally, the use
of RISC-V in our platform makes extending the ISA easier, given RISC-V's opcodes
dedicated to user defined instructions \cite{risc_v_spec}. These improvements
in our platform create a richer and larger explorable design space.

The Soft Processor Rapid Exploration Environment (SPREE) tool provides another
MIPS-based design space exploration tool \cite{spree2007} \cite{spree2006}.
SPREE explores such trade-offs in
micro-architectural details as pipeline depth, hazard detection implementation, as well as ISA
features like branch delay slots and application-specific register management.
However, the design space covered by SPREE is limited to the cores in a processing system.
Additionally, only single core designs are supported by SPREE.
Our platform offers a more complete design space exploration, with support
for multi-core systems with core, cache, memory and NoC configuration options.

The free and open nature of RISC-V means that numerous open-source implementations
are available. A few of these, for example, a size optimized core named PicoRV32
\cite{wolf2018picorv32} and a Linux capable core named RV12 from RoaLogic \cite{rv12},  include SoCs generated with the Rocket
Chip Generator \cite{asanovic2016rocket}.
Many more implementations exist, but for brevity they are omitted.
While the RISC-V implementations mentioned offer some level of configuration,
none of them support multi-core, cache, or NoC design space exploration in the
way our platform does. Table~\ref{comparison} compares the configuration options and features available for design space exploration
in each of the referenced works.

\vspace{-0.1in}
\begin{center}
  \begin{table}[h]
    \begin{tabular} {  |r |g|w|g|w|g|w|  }
      \multicolumn{1}{c}{}& \rot{PicoRV32} & \rot{RV12} & \rot{Rocket Chip} & \rot{SPREE} & \rot{Heracles} & \rot{Our Platform} \\
      \hline
      Core                  & \greencheck & \greencheck & \greencheck & \greencheck & \greencheck &  \greencheck \\
      \hline
      Cache                 & & \greencheck & \greencheck & & \greencheck  & \greencheck  \\
      \hline
      Memory                & & & & & \greencheck  & \greencheck  \\
      \hline
      Interconnect          & & & & & \greencheck  & \greencheck  \\
      \hline
      Multi-Core            & & & \greencheck & & \greencheck & \greencheck \\
      \hline
      Extensible ISA        & \greencheck & \greencheck & \greencheck & & & \greencheck \\
      \hline
      GUI                   & & & & & \greencheck & \greencheck \\
      \hline
    \end{tabular}
    \caption{A comparison of configurable subsystems and features available in
    popular configurable processors.}
    \label{comparison}
    \vspace{-0.3in}
  \end{table}
\end{center}

\vspace{-0.15in}
\section{Platform Overview}
\begin{figure}[t]
    \centering
    \vspace{-0.05in}
    \includegraphics[width=1.00\columnwidth]{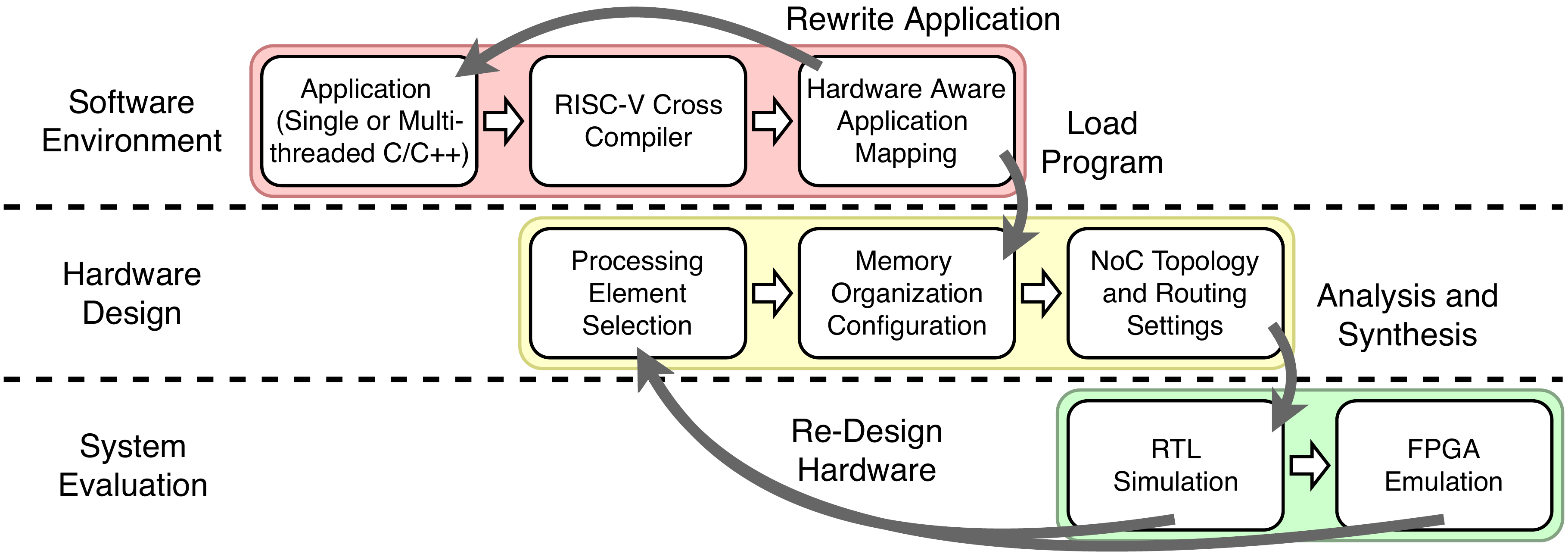}
    \vspace{-0.2in}
    \caption{A typical workflow for the platform.}
    \label{flow}
    \vspace{-0.15in}
\end{figure}

A typical workflow in the platform (shown in Figure~\ref{flow}) starts
with developing the software application to be run on the hardware
system.  Software is compiled with either the GNU or LLVM
  compiler toolchain for RISC-V. The included compiler scripts
support simple multi-processing and multi-threading environments.

After developing the application, users can determine performance requirements
as well as power and area constraints. Given the system requirements, a user
can begin setting parameters for the core, cache, memory and NoC subsystems.
Users can set these parameters with the hardware configuration GUI discussed in
Section~\ref{section:gui}. Users requiring a small processor
to handle embedded applications might select a small single cycle core with a
simple BRAM memory, while a user developing a large distributed multi-core
system could opt for the more complex pipelined or out-of-order cores, with
large caches and a memory controller for off-chip memory.
Users developing a system for a single application can optimize
the cache subsystem by selecting line size, associativity, and number of line values
best suited to the memory access pattern of the application.
For instance,
if the application has high spatial locality, then a larger line size can be selected.
If users are designing a large many-core processor they can experiment with different NoC
topologies and routing algorithms to determine which one gives the best performance for their
constraints.

The test benches included in the hardware system can be used to simulate the processing
system and verify that the user's program executes correctly. Tests for the
sample programs included in the toolbox have been automated to report a pass or
fail result to accelerate development of custom features. After passing
simulation tests, the hardware system can be synthesized for implementation on
an FPGA. The ability to generate a processing system rapidly allows users to iterate
their design in the event that their requirements are not met by the initial
system. At any point in the design flow, users can easily go back and tweak the design
to meet their constraints better.
Quickly iterating a design enables users to
develop the hardware and software systems together,
facilitating a thorough design space exploration.

\section{Platform Core Descriptions}
\label{core_description}
A single cycle processor is presented as a baseline for processor design,
emulation, and analysis. For users to explore the impact pipeline depth has on
performance and area, the platform includes five- and seven-cycle pipelined processors.
For users to explore instruction extensions and a wide variety of
micro-architecture features, the platform includes a super-scalar out-of-order
processor.

\subsection{Single Cycle Processor}
\begin{figure}[h]
   \vspace{-0.05in}
    \centering
    \includegraphics[width=1.0\columnwidth]{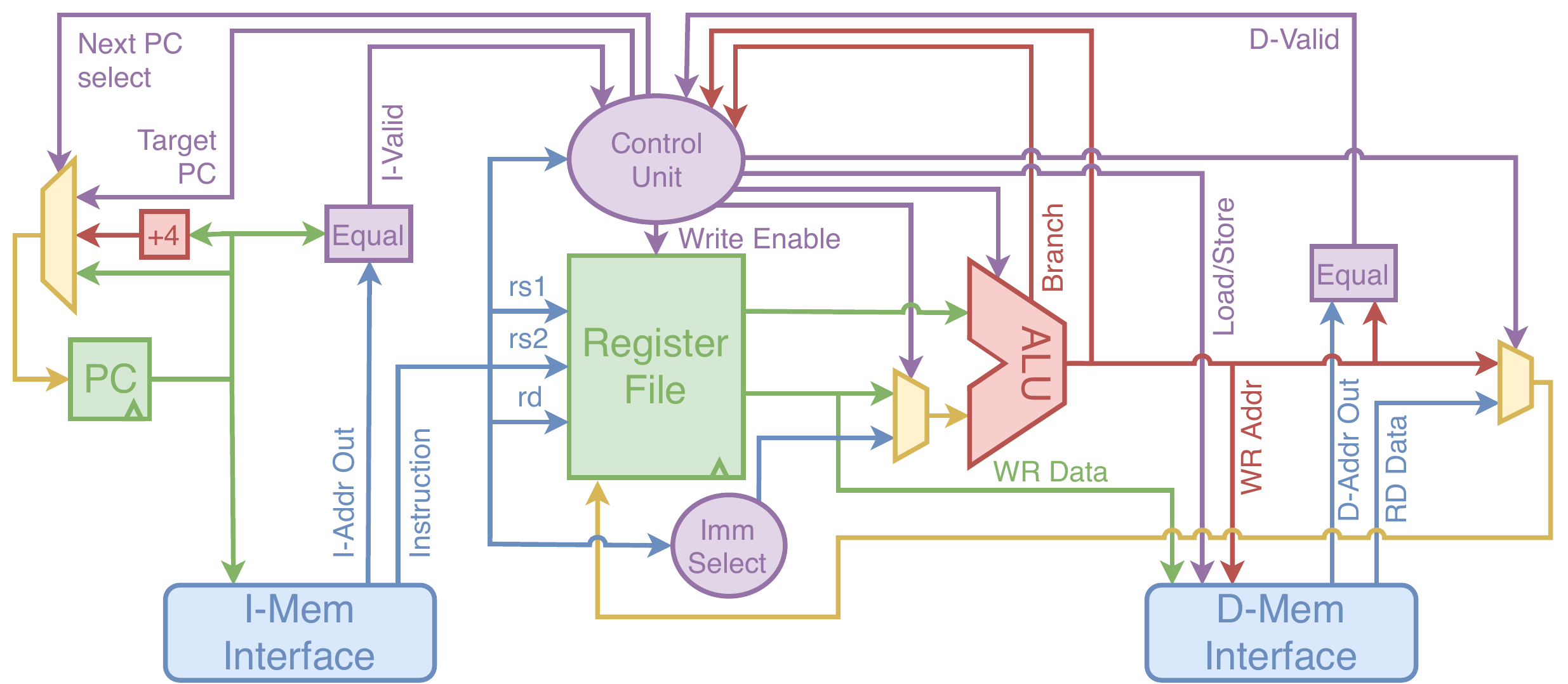}
    \vspace{-0.2in}
    \caption{RV32I single cycle core.}
    \label{single_cycle}
    \vspace{-0.15in}
\end{figure}

The single cycle processor implements the RV32I instruction set with modules 
designed  around the ``textbook'' fetch, decode, execute, memory, and write-back
stages of a processor \cite{Patterson:2013:COD:2568134}. 
This processor serves as the base for other cores; as such, it has been designed
to be as simple as possible. The modules in the single cycle processor are
reused or wrapped with additional logic to support features such as pipelining
and data forwarding.
A block diagram of the
processor is shown in Figure~\ref{single_cycle}.
The single cycle processor has instruction and data memory interfaces
compatible with every cache and memory subsystem provided in the platform.
Due to the single cycle operation, NOPs are inserted between BRAM or off-chip
memory accesses. An asynchronous memory is provided to avoid NOPs, but it
cannot be implemented in FPGA BRAM and must be kept small to prevent the memory
from using too many device resources.

By using the Verilog hex output from the included compiler toolchain
described in Section \ref{compiler},
users are able to compile a bare-metal C program and run it on an FPGA
implementation of the processor. Synthesis results have been collected and are
shown in Table~\ref{synthesis_results}.
Note that the Logic Element usage is
high because the memory is implemented in Look Up Tables (LUT), because of the
asynchronous memory system used.

\subsection{Five Stage Pipeline}
The five stage processor is implemented by using the base modules from the single
cycle processor as a starting point and adding pipeline registers between the
combinational fetch, decode, execute, memory and writeback modules. The single cycle control logic module is wrapped with additional logic to support the stall and bypass signals needed for pipelining. Introducing multiple instructions in flight demonstrates how hazard resolution must consider bypassing, stalling, and pipeline flushing.

Pipelining allows for a higher clock frequency; however, NOPs are still inserted
between synchronous memory operations because there is no pipeline
register between the address input and data output of the memory interfaces.
Asynchronous memories must still be used to avoid NOPs between instruction
fetches.
Note, however, that the addition of pipeline stages allows some synthesis tools
to implement an asynchronous main memory (without caches) in FPGA BRAM.
The five-stage pipelined processor has two variants. The first uses only pipeline
stalls and flushes  when a pipeline hazard is detected. The second implements
data forwarding to avoid stalls for most hazards.
Stalling and forwarding logic is wrapped around the base control unit used in
the single cycle core. A multiplexer is wrapped around the decode logic to
output forwarded data when needed.
Wrapping the base modules to build the five stage pipeline maximizes IP reuse and
allows for user base module customizations to be carried through
their core design space.

\begin{figure}[h]
    \centering
    \includegraphics[width=1.0\columnwidth]{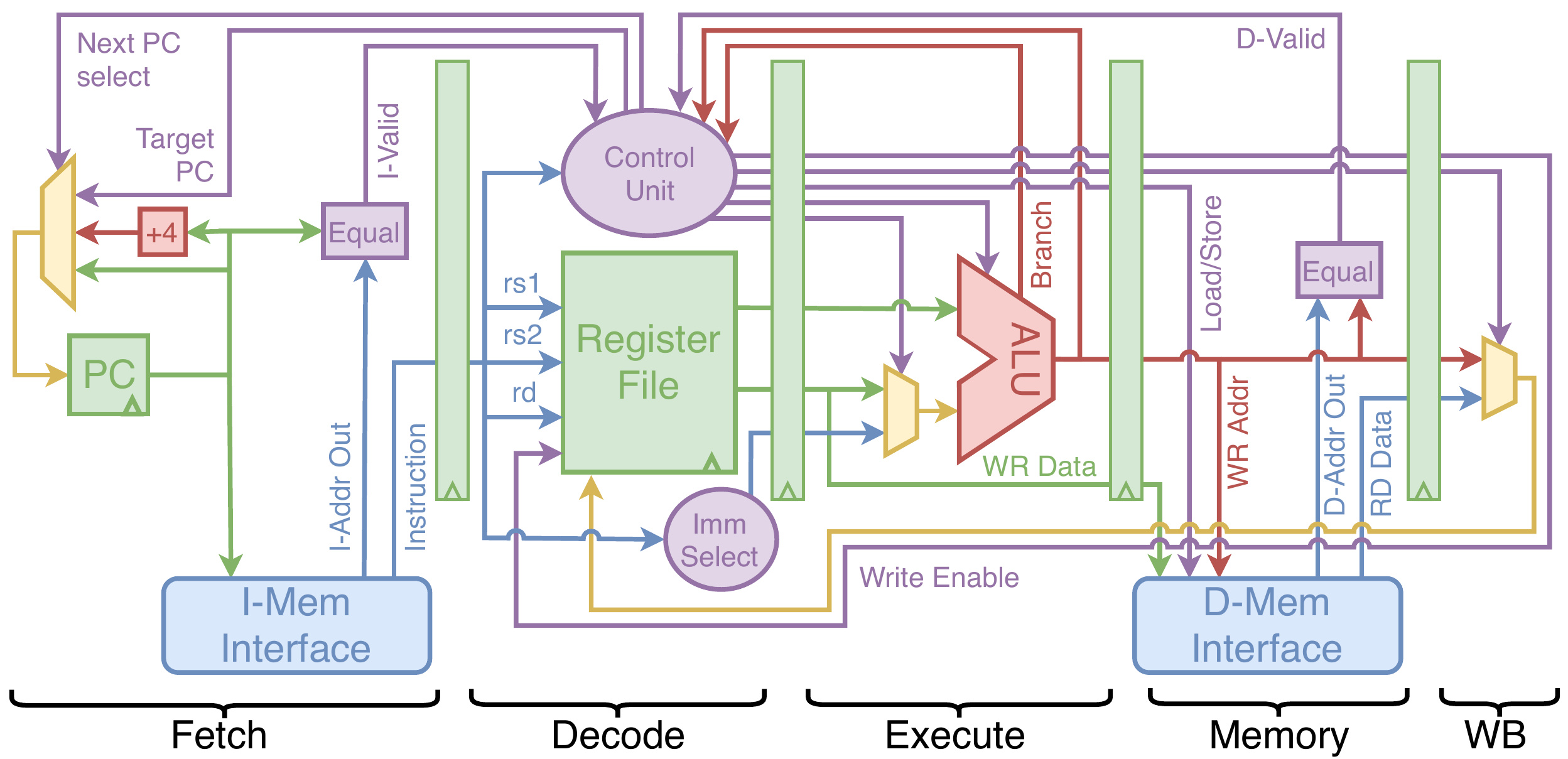}
    \vspace{-0.2in}
    \caption{RV32I five-stage pipelined core.}
    \label{pipelined}
    \vspace{-0.10in}
\end{figure}

\subsection{Seven Stage Pipeline}
\begin{figure}[h]
    \centering
    \includegraphics[width=1.0\columnwidth]{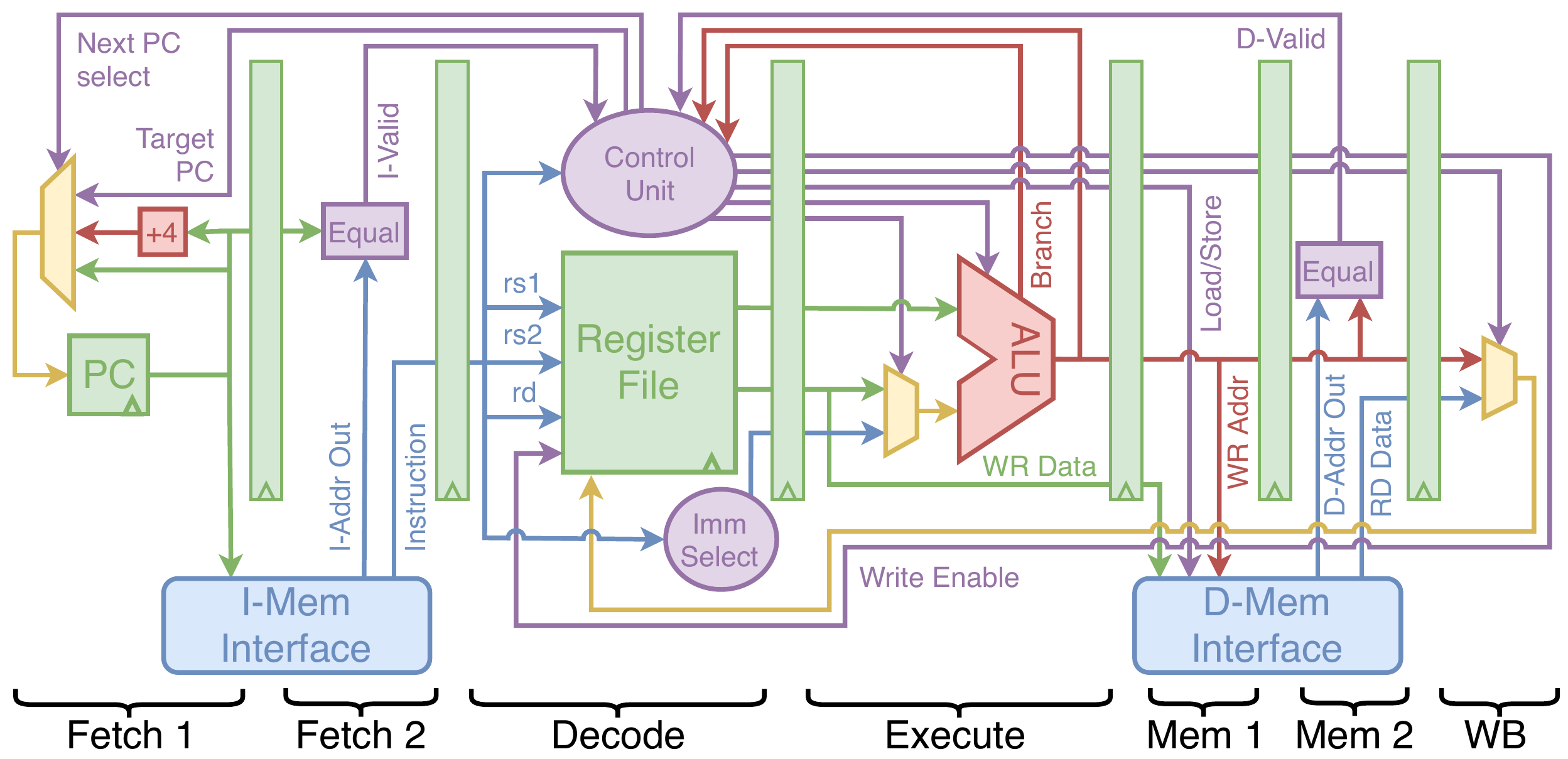}
    \vspace{-0.2in}
    \caption{RV32I seven-stage pipelined core.}
    \label{seven_stage_pipeline}
    \vspace{-0.10in}
\end{figure}

The seven-stage pipelined processor builds on the base modules in the
single cycle processor and the pipeline related modules introduced with the
five stage processor. It adds registers between the address inputs and data outputs
of the memory interfaces to avoid inserting NOPs while waiting for synchronous
memory operations. With these extra pipeline stages, BRAM reads and cache hits
no longer need to insert NOPs. The additional pipeline stages can be seen in the seven stage pipeline block
diagram shown in Figure~\ref{seven_stage_pipeline}.

Placing extra stages between the memory interface input and outputs enables
logic to check that a memory read is valid while the next operation is
issued. Operations in cacheless implementations with on-chip BRAM will always be valid;
however, the addition of caches means that cache misses could delay valid read
data. In the event of a cache miss, the received
data is marked invalid by the memory and the processor stalls until the requested memory has
been retrieved. On cache hits, execution continues normally with no inserted
NOPs. The extra registers in the seven stage pipeline yield an improved maximum clock
frequency. The extra pipeline registers lead to simplified control logic,
resulting in a slightly reduced area.
Synthesis results for the seven stage pipeline are shown in Table~\ref{synthesis_results}.

\subsection{Out of Order Processor}
The Out-Of-Order (OOO) core enables exploration of advanced architectural
features including a superscalar architecture, instruction scheduling, and
complex hazard resolution. The OOO processor supports out-of-order execution
with in order commit. The number of ALUs has been parameterized to allow users
to explore the impact of a varying number of functional units on processor
performance.

The OOO processor implements the RV32F instruction set extension
to create more opportunities for out-of-order execution.
In order to support the floating point extension, a floating point register file
and floating-point execution units were added to the processor.
The base decode and control units have been expanded with extra logic to support
the new floating point instructions.

\begin{figure}[t]
    \centering
    \includegraphics[width=1.00\columnwidth]{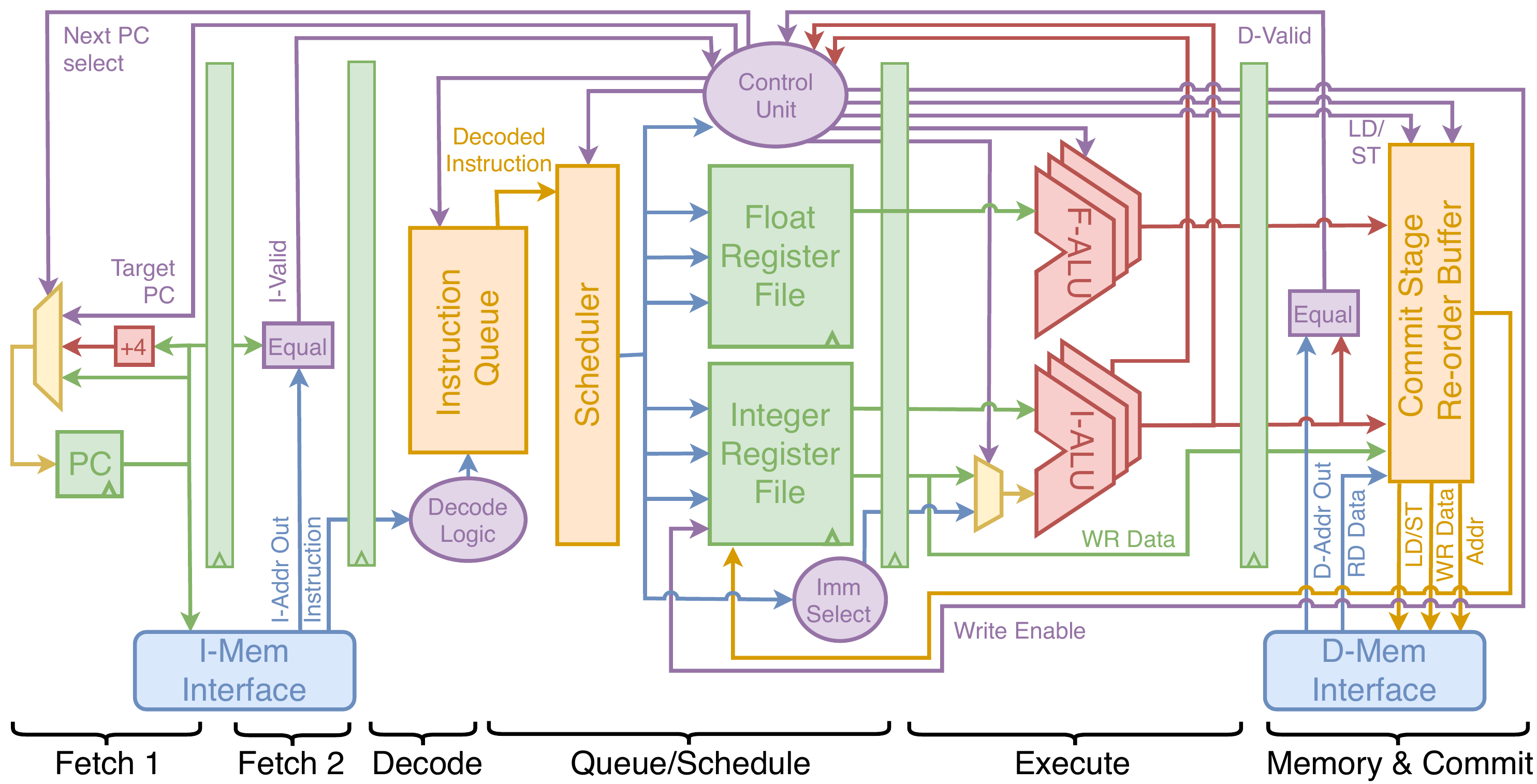}
    \vspace{-0.2in}
    \caption{RV32IF Out-of-order core.}
    \label{oooe}
    \vspace{-0.3in}
\end{figure}

The OOO core adds three multi-cycle modules to the processor pipeline: (1) an
instruction queue, (2) a scheduler, and (3) a commit stage. These three
stages do not reuse any of the base modules from the in-order cores discussed
previously. Figure~\ref{oooe} shows a block diagram of the OOO micro-architecture.

In the out-of-order processor, instructions are fetched and decoded before being
placed in the new instruction queue stage. The instruction queue length can be
modified by the user to trade off performance and area. The queue is implemented
as a priority queue, in which the highest priority (longest waiting) instruction
without any hazards is scheduled next.

The scheduler module supports a parameterized number of floating point and 
integer ALUs. When both an instruction and an ALU are ready, the scheduler assigns
the instruction to the available ALU.
After the instruction has completed its execution with respect to the ALU, it
continues through the pipeline into the commit stage. Up to one instruction
is scheduled each clock cycle. ALUs can be pipelined to ensure their utilization
remains high, offering yet another design space exploration trade off.

Memory operations wait in the commit module until the memory access has completed.
Instructions in the commit module are buffered and sent to the writeback unit
in order. Instructions are re-ordered based on the priorities used in the
instruction queue.
An instruction's `rd' value is stored in a table in the instruction queue stage.
Hazards are detected by comparing the `rd' values of in-flight instructions
to the `rs1' and `rs2' values of instructions entering the queue. An instruction's `rd'
value is cleared from the table in the instruction queue stage after the instruction exits 
the writeback stage, completing its execution.

\section{Cache subsystem}
Our platform includes a highly configurable multi-level cache subsystem in order
to provide a high degree of freedom to users.
Different cache configurations can be implemented by adjusting parameters.
The cache subsystem can be
easily modified to fit different performance requirements or available resources. For instance,
if the system is to be implemented on a smaller FPGA, one or two levels of smaller caches
could be used. Alternatively, if a larger FPGA is available, larger caches and more
levels in the cache hierarchy would provide better performance.  The ability to
test a large number of different cache configurations, without investing time to
develop the different cache systems from scratch, streamlines cache-focused
design space exploration.

The cache subsystem supports multi-stage inclusive caches. The caching policy is
``write back with write allocate''.
Currently, the cache system supports ``MESI'' cache coherence. Other system parameters, such as
the number of cache levels, cache size, and internal parameters of each cache are configurable.
Coherent cache configurations with heterogeneous line widths and associativities are also supported.
The cache subsystem is comprised of two fully parameterized cache modules, a shared bus, and a
coherence controller.
An interface module is included to act as the interface between the last level cache and the
main memory or on-chip network.

\begin{figure}[b]
    \centering
    \vspace{-0.05in}
    \includegraphics[width=0.95\columnwidth]{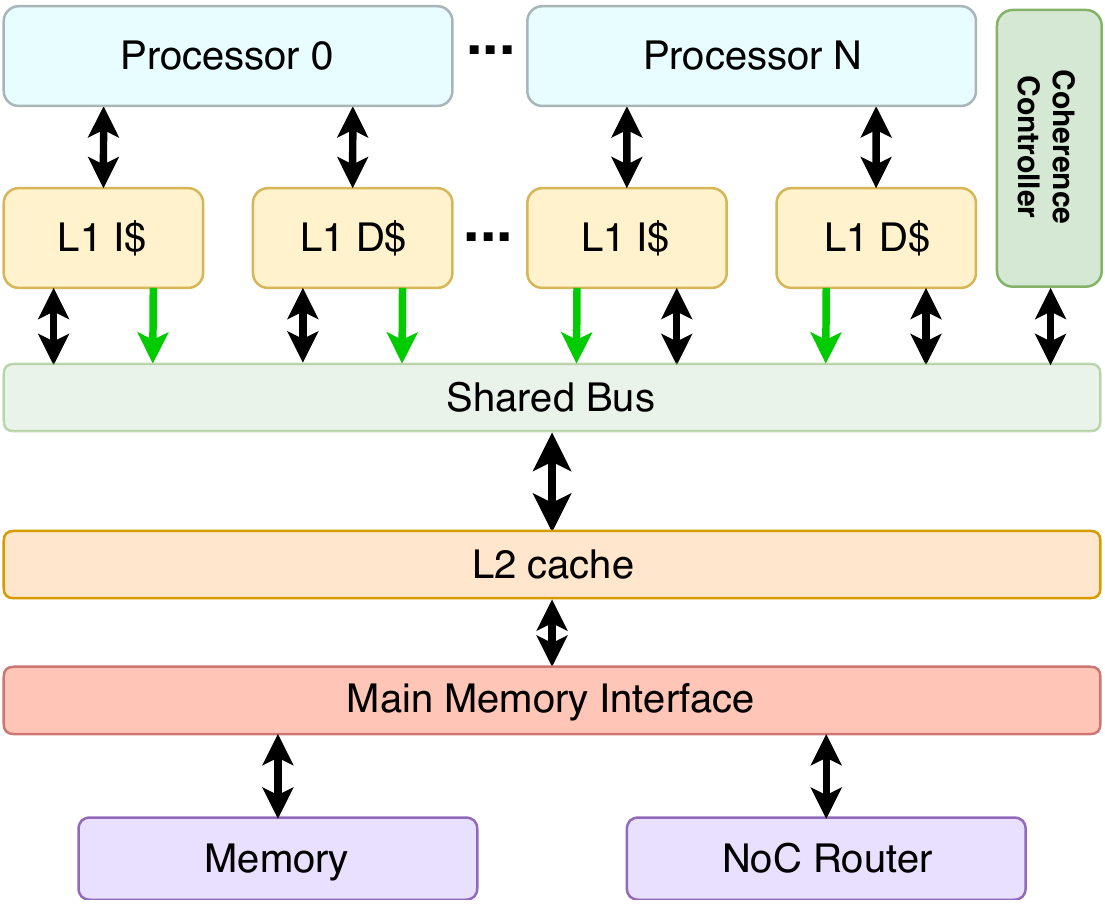}
    \vspace{-0.1in}
    \caption{Multi-core memory hierarchy block diagram.}
    \label{cache}
    \vspace{-0.1in}
\end{figure}

The ``main\_memory\_interface'' module bridges the gap between main memory word
size and last level cache line size.
This decoupling allows the user to use off-chip memory as the main memory. In the case of a
distributed memory system, this module also acts as the interface between the last level
cache and the on-chip network. Figure~\ref{cache} depicts the platform's cache
hierarchy.

\subsection{L1cache}
The ``L1cache'' module is designed to be used as the level 1 cache that interfaces with a
processor. It provides user configurable parameters for cache size, cache line width, associativity,
and replacement policy. The ``INDEX\_BITS'' parameter determines the number of cache sets.
``OFFSET\_BITS'' parameter determines the cache line width while ``NUMBER\_OF\_WAYS'' parameter
specifies the associativity of the cache. RTL for the cache is written in a highly parameterized
fashion with additional parameters for data width, address width, number of status bits, number of coherence
bits, number of bits to select replacement policy, and number of bits used for communication with
the other levels of the cache hierarchy.
These parameters make it simpler to build extensions to the cache subsystem by
minimizing the number of RTL changes required.

The ``L1cache'' receives six signals from the processor. The processor specifies
the memory address to access with the ``address\_in'' port and the data to write
with the ``data\_in'' port.
Additional ports exist for the read, write, flush, and invalidate signals. Flush and invalidate
operations are carried out on a cache line granularity. When the processor specifies a single
memory address with the flush/invalidate signal, the cache line containing the address is \\
flushed/invalidated throughout the entire cache hierarchy.
If the cache line was dirty, it is written back to lower levels in the cache hierarchy
and ultimately to main memory. Since the caches are inclusive, a flushed cache line
is also flushed/invalidated from L1 caches of other cores in the system.

The L1 cache sends four signals to the processor. The ``data\_out'' bus sends the data read from
memory, while the ``out\_address'' is the memory address corresponding to the data on the
data bus. The ``valid'' signal indicates that the data on the data bus is valid. The ``ready''
signal informs the processor that the cache is ready for the next memory request. The
processor should stall parts of the pipeline based on ``valid'' and ``ready'' signals.
Both signals are required to convey the cache status to the processor
because the L1 cache operates in a pipelined fashion with up to two accesses in
flight at any time. The L1 cache has a one-cycle access time due to FPGA
BRAM access latency.
Each way of the cache is mapped to a separate BRAM on the FPGA.

Both ``L1cache'' and ``Lxcache'' modules use the same interface for communicating
with the lower levels (caches more distant from the processor) in the cache
hierarchy. A common interface is used to enable an arbitrary number of cache
levels in a design. This interface consists of six signals:
data\_in, address\_in, message\_in, data\_out, address\_out and message\_out.

Data in/out contains a whole cache line. All caches use different 4-bit messages
to communicate with other caches via the shared bus.
There are two types of messages. The first type is to communicate basic requests such as
read, write back, flush, etc..
The second type of messages is related to cache coherence. When an L1 cache is writing to
a shared line or reading a cache line on a write miss, it broadcasts its intent so that
the other caches can perform necessary coherence operations. L2 cache issues flush
requests to L1 caches when evicting a cache line that one or more L1 caches have a copy of.
Cache coherence is discussed in detail in Section~\ref{coherence}.

Currently, the caches can be configured to perform true Least Recently Used (LRU) or
random replacement of cache lines. Because of the modular design of the ``replacement\_controller''
module, a user can easily implement other replacement policies.

\subsection{Lxcache Module}
The ``Lxcache'' module is configurable to be used at any level in the cache
hierarchy except for level 1 where the cache interfaces with the processor.
``Lxcache'' supports all the configurable parameters in ``L1cache''.
The ``Lxcache'' module adds the capability to serve an arbitrary number of ports
with round robin arbitration. Multiple ports enable level 1 caches from several
processors to be connected to a shared level 2 cache.

\subsection{Cache coherence}
\label{coherence}
The platform's cache subsystem implements ``MESI'' cache coherence which makes it possible to build
and test multi-core architectures. The ``coherence controller'' module is designed to be
instantiated alongside L1 and L2 caches. This module is capable of serving an
arbitrary number of L1 caches and the L2 cache. The shared bus between L1 and L2
caches is controlled by the coherence controller. It listens to messages
issued by L1 and L2 caches and controls which cache drives the shared bus.

The L1 caches are designed to listen and respond to messages on the shared bus by
either writing back or invalidating cache lines.
Different operations, of which a read request is the simplest, trigger cache coherence operations. 
If one of the other caches has a dirty copy (in ``MODIFIED'' state) of the requested cache line,
it will write back the dirty cache line to the shared bus.
The coherence controller will allow the write back to go on the bus so that the L1
cache that issued the original read request and the L2 cache can update the
cache line in question.
If any of the caches have ``exclusive'' copies of the cache line,
they will be changed to the ``SHARED'' state.

When an L1 cache is writing to a shared cache line,
it sends a message communicating its action. That message is broadcasted to all
other L1 caches over the shared bus. Upon receiving the message, other caches invalidate
the shared line and respond to the coherence controller. Once all caches respond to the
broadcast, the coherence controller puts ``NO\_REQ'' message on the bus. This indicates
to the first cache that it can write to the shared line and change its coherence status to
``MODIFIED''.

Another type of coherence operation is triggered when an L1 cache encounters
a write miss. Since it intends to write to the cache line read from the lower level, the
L1 cache sends a message requesting ownership of the cache line instead of a standard read
request. This indicates to the other L1 caches that unlike with a read request, they should
invalidate the cache line if they have copies of it in ``SHARED'' or ``EXCLUSIVE'' states. 

The last type of coherence operation occurs when the L2 cache is about to evict a cache line
that is also in one or more of the L1 caches. Since this is an inclusive cache hierarchy,
the copies of the cache line should be evicted form L1 caches as well. Therefore, the
L2 cache issues a flush request to the L1 caches.The  L1 caches either invalidate or write-back the
cache line depending on its status.

L1 caches are designed with an independent snooper module to perform 
coherence operations. Dual ported block RAMs are used as memory to perform coherence
operations without interrupting normal memory accesses by the cache controller.
When both the cache controller and the snooper access the same cache 
line, the coherence operation is given priority.
Bus interface also gives the snooper priority when both the controller and snooper 
attempt to access the shared bus.

\subsection{Limitations}
Currently, cache coherence is handled at the first level of the cache hierarchy (L1 caches).
This forces the L2 to be shared between all processors. Future improvements to the cache
subsystem will allow the user to choose whether cache coherence is handled at L1 or L2 level.
Moving the coherence controller to the L2 level will allow a user to build a cache
hierarchy with private L1 and L2 caches, and a shared L3 cache.

\section{Main memory and Network-on-chip}

The main memory interface decouples the cache subsystem and the main memory.
Users have the option to use any of the provided main memory
subsystems: (1) unified or separate asynchronous instruction and data memory,
(2) unified or separate synchronous instruction and data memory, or (3) off-chip memory controller.

Connecting the main memory interface to an off-chip memory controller enables
large main memories. Using off-chip memory is useful when a system needs more
memory than is available on a given FPGA. Currently the platform includes a simple
off-chip SRAM memory controller; other device specific memory controllers can
easily be added.
The interface also supports connections
to an on-chip network, which, coupled with the ability to configure the size of
main memory on a per node basis, enables uniform and non-uniform distributed
memory systems.

\section{On-Chip Network}
\begin{figure}[htb]
    \centering
    \vspace{-0.1in}
    \includegraphics[width=0.8\columnwidth]{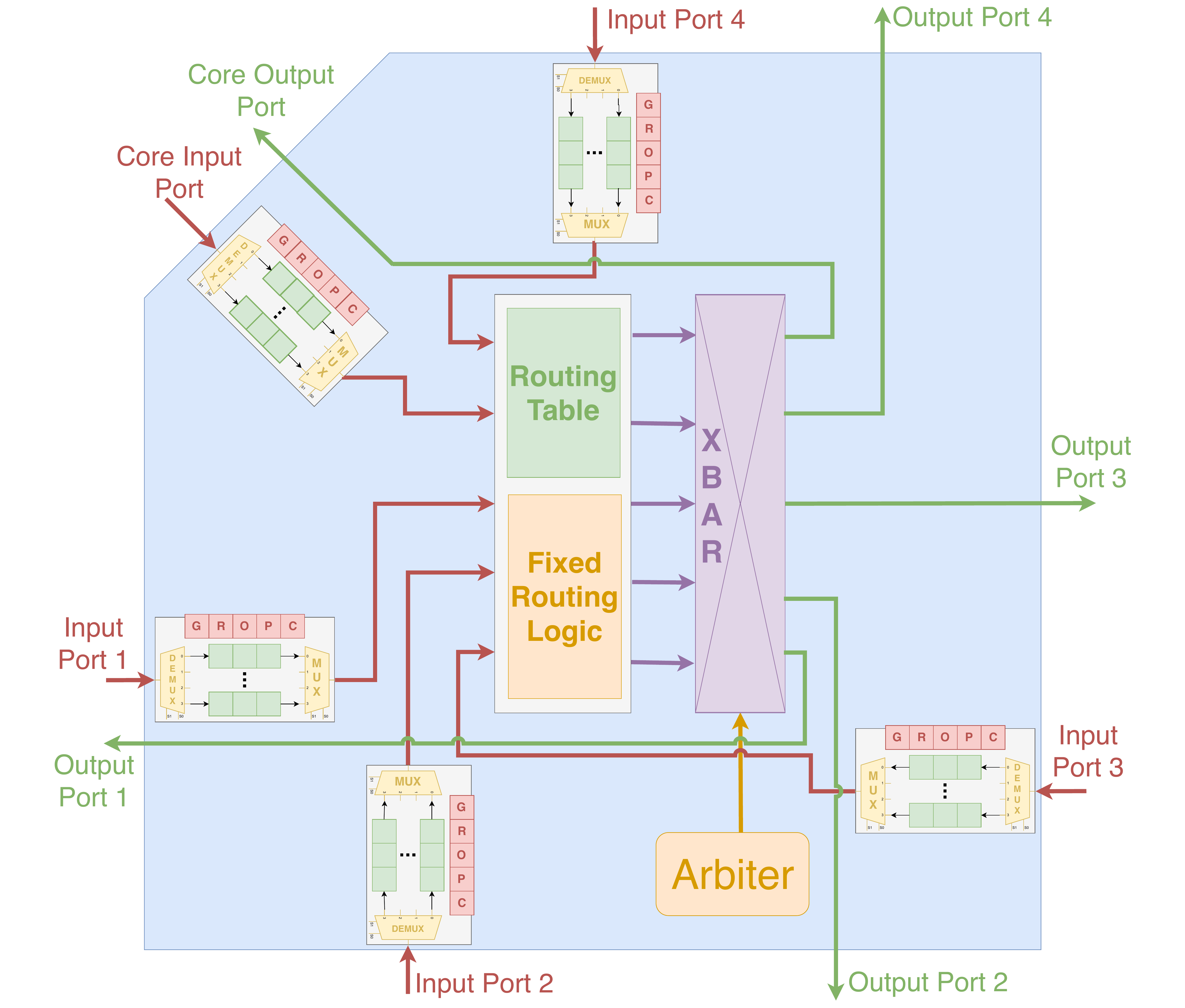}
    \vspace{-0.1in}
    \caption{NoC router architecture.}
    \label{router_fig}
\end{figure}

The platform's on-chip network works with the memory subsystem to implement
a variety of multi-core architectures. The NoC provides a number of
configuration options, enabling the user to explore different network
topologies and optimize the resource usage and performance of the system.
The network can be configured to explore different combinations of: 1)
flow control,  2) routing algorithms, and 3) network topologies. The NoC router is fully parameterized~\ref{router_fig}. 

Routers in the NoC can be configured as buffered or buffer-less routers. The
routers support oblivious routing algorithms using fixed logic or configurable
routing tables. Fixed logic is implemented for Dimension Order Routing.
Programmable routing tables enable different routing algorithms with changes to 
the routing table entries.
A wide range of network topologies can be be implemented by configuring the
number of input ports, output ports and routing table contents of the routers.
Routers included in the platform are conventional virtual channel routers.
There are single cycle and pipelined variants of the routers.
Users can configure different parameters such as number of input/output ports, 
virtual channels per port, virtual channel depth to tune the performance and 
resource usage of the on-chip network opening a rich design space for exploration.
The on-chip network is based on the NoC included with the Heracles system
\cite{kinsy2011heracles} \cite{kinsy2013heracles}.

 \vspace{-0.15in}

\section{Workflow Description}
Being able to make changes to the hardware system easily and
understand the performance implications of those changes make our platform a
powerful micro-architecture design space exploration tool. Quick design changes
are supported with open-source parameterized Verilog modules and a hardware
system configuration GUI.
The configuration GUI provides a user friendly way to choose parameters and
visualize a hardware system.  A compiler toolchain streamlines software
development. A RISC-V GCC cross-compiler binary is included, so users do
not have to configure and build the RISC-V tools from source code. The following
subsections describe the workflow for the software toolchain and hardware
configuration GUI.

\subsection{Hardware Configuration GUI}
\label{section:gui}
The hardware system configuration application is a graphical application that allows
users to configure a hardware system to meet their specification. The
application runs in a web browser, allowing users to run it on Windows,
Linux, or Mac.

With the hardware configuration system users can (1) select their desired core
type and features; (2)  include a cache subsystem, if desired, and select its
parameters; (3) choose a main memory subsystem, e.g. on-chip, off-chip, unified
or separate instruction and data memories; and (4) choose NoC configuration
options including number of routers, router topology, and router scheme.

\begin{figure}[t]
    \centering
    \vspace{-0.1in}
    \includegraphics[width=0.95\columnwidth]{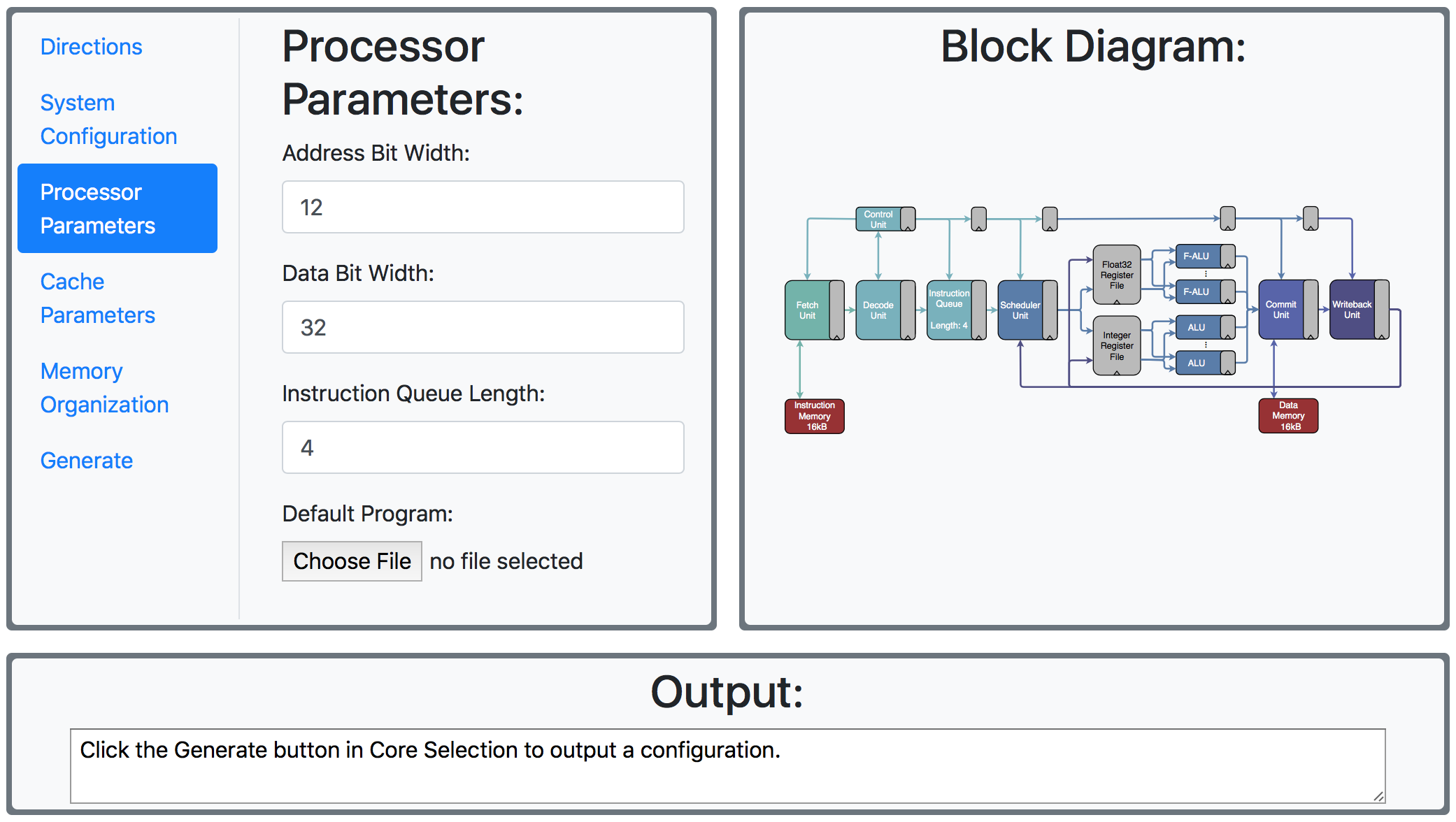}
    \vspace{-0.1in}
    \caption{A screenshot of the hardware configuration GUI. Note that logos have been
    cropped out to maintain anonymity.}
    \label{brisc_v_explorer}
    \vspace{-0.10in}
\end{figure}

Figure~\ref{brisc_v_explorer} shows a
screenshot of the application.
On the left are menu and
parameter entry text boxes. A block diagram of the configured system is
shown on the right of the application window. 
Selecting different menu options
opens different parameter selection tabs.
The five core types currently included are 1) single cycle, 2) five-stage pipeline with
stall on hazard, 3) five-stage pipeline with data forwarding, 4) seven-stage
pipeline with forwarding, and 5) pipelined Out-Of-Order.
As the processor cores gain complexity, so do their parameters.
The cores are built off of one another, with each core serving as the starting
point for the next more complex core.
Each new implementation supports the previous processor's parameters, in addition
to any new required parameters. If on-chip main memory is selected, it can be initialized with the ``PROGRAM''
parameter. This parameter points to a Verilog Memory Hex file that is output by
the provided compiler toolchain described in Section~\ref{compiler}.

After a user has configured the system, clicking ``Generate and download'' will
download the configured RTL and selected binary from the application. Note that
the application is run in a web browser but the entire application can also be local to
a user's machine. No internet connection is required to use the configuration
GUI. The use of ``Download'' here refers to the fact that the browser is oblivious
to the application's origin.
Running the hardware configuration GUI in a browser enables researchers,
students and teachers to use it locally on a platform of their choice or host it
on a server for users to access remotely. Hosting the configuration GUI on a
server could simplify its use in a classroom environment.

\subsection{Compilation}
\label{compiler}
Software for the platform can be built using the standard
GNU~\cite{riscv_tools} or LLVM~\cite{llvm_tools} compiler toolchains.
The GNU toolchain is the default and is distributed in the binary form with
the project. The binary distribution includes compiler, assembler,
linker and the standard library. For users who decide to use LLVM
toolchain, we distribute detailed instructions for building a
bare-metal RISC-V LLVM backend. RISC-V is supported as an experimental
target from LLVM version \texttt{8.0.0} and can easily be enabled
during the build process. Beside support for modern
programming languages such as Rust, LLVM infrastructure provides
modular mechanisms for adding custom instructions and compiler
optimizations. The educational material that covers the writing of
compiler backends and custom optimizations in the form of ``passes''
is publicly available. These features of the LLVM infrastructure are
beneficial for design space exploration.

A script is included to compile user code and convert it to a
format that can be synthesized as ROM or initialized RAM for
implementation on an FPGA.
The provided compilation script outputs (among other formats) an ASCII
encoded Verilog Memory Hex (\texttt{.vmh}) file. This \texttt{.vmh}
file can be used to initialize memory contents on an FPGA with the
Verilog \texttt{\$readmemh()} function.

\begin{figure}[h]
  \centering
   \includegraphics[width=1.00\columnwidth]{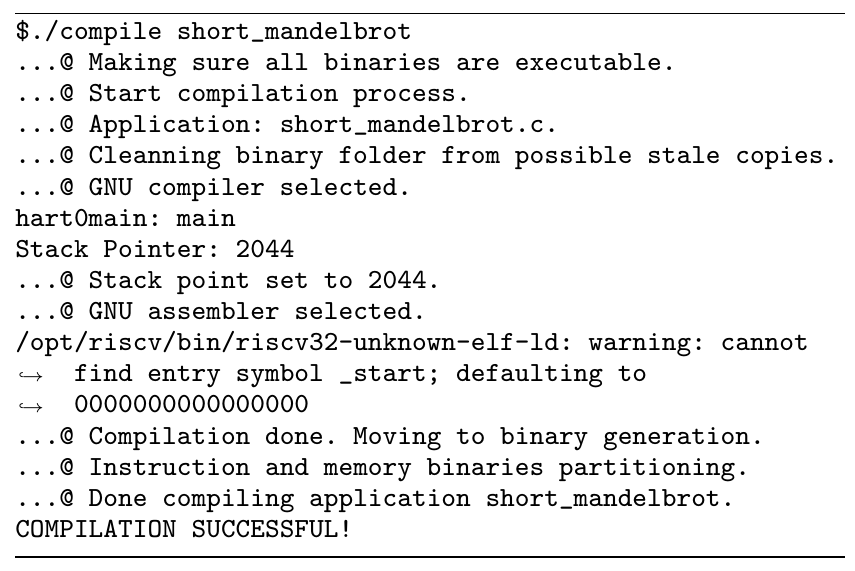}
\vspace{-0.25in}
\caption{Example Shell command to compile a program named
  short\_mandelbrot.c with an initial stack pointer of 2044 (0x7fc).}
  \label{fig:compilation}
    \vspace{-0.1in}
\end{figure}

The provided compilation script can be found in the ``software''
directory. Application source code should be placed in the
``software/applications/src'' directory. In order to compile an
application, the user runs the ``compile'' script from the software
directory. Figure~\ref{fig:compilation} shows the command usage and
output.

The first argument to the compile script is the name of the C file
that user wants to compile. The remaining arguments are optional. The
second argument sets the initial stack pointer. The third argument
sets the size of the stack given each hardware thread (HART) and a
fourth argument specifies the number of harts in the system. By
default the stack pointer is 2044 (0x7fc) and only one HART is
used. The stack pointer may need to be changed if the selected memory
size is very small. The compilation script places the resulting
\texttt{.vmh} file in the ``software/applications/binaries/''
directory alongside with the other outputs.

The project currently does not support any operating system and
applications must be run in a bare-metal environment. To eliminate any
extra setup by the user to enable bare-metal execution of their code,
the compiler script wraps the user application with assembly code to
initialize each of the registers and call the main function. The
assembly wrapper executes an infinite loop after the main function
returns. This wrapper can be thought of as a simple kernel.
Multiple HARTs can be used by writing a function named \texttt{hartN\_main()}
for each HART, where ``N'' is the HART number. The simple kernel will call each
HART's main function.

The user's custom compiler choice as well as the compiler options
should be provided as environment variables. The environment variable
used to supply additional options to GCC compiler is
``GCC\_OPTS''. The variables to control the version of LLVM compiler
and the additional options are ``LLVM'' and ``LLVM\_OPTS'',
respectively. The user should provide these options before they run
the compilation script. That being so, we maintain the flexibility of
the compilation process and support the standard tools for code
optimizations.

After running the provided compile script, a user can set the
``PROGRAM'' parameter to the path of their \texttt{.vmh} file
(manually or in the hardware configuration GUI) and synthesize their design. The
instruction memory of the processor will be initialized with the
contents of the binary if it is on-chip. The instruction memory also has a port for
writing new programs after the design has been configured on an FPGA.

\section{Platform Evaluation}

\subsection{Stress Testing Example of the Platform}
\begin{figure}[t] 
  \subfigure[]{%
    \includegraphics[width=.48\columnwidth,height=0.1\textheight]{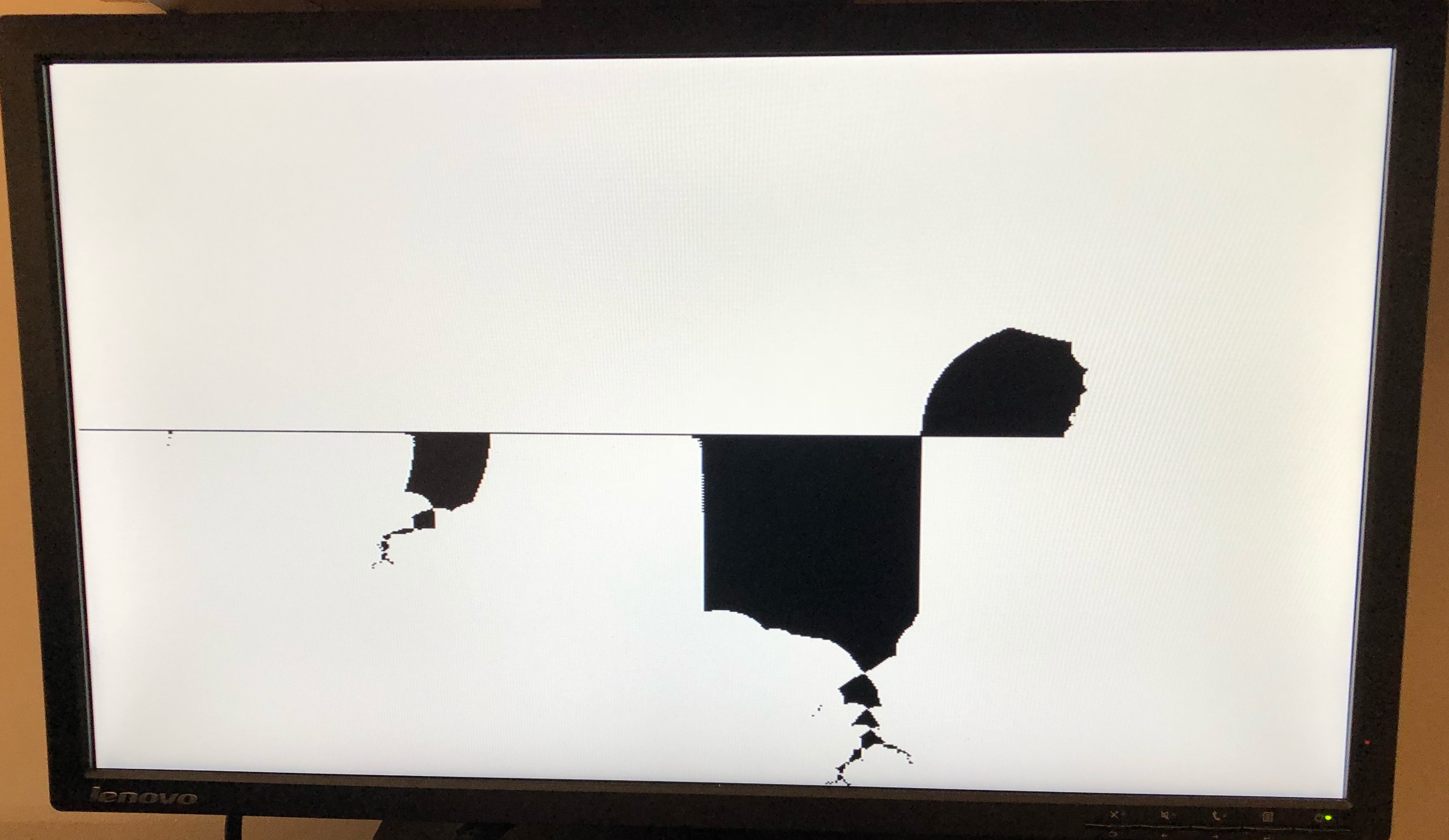} 
  } 
  \subfigure[]{%
    \includegraphics[width=.48\columnwidth,height=0.1\textheight]{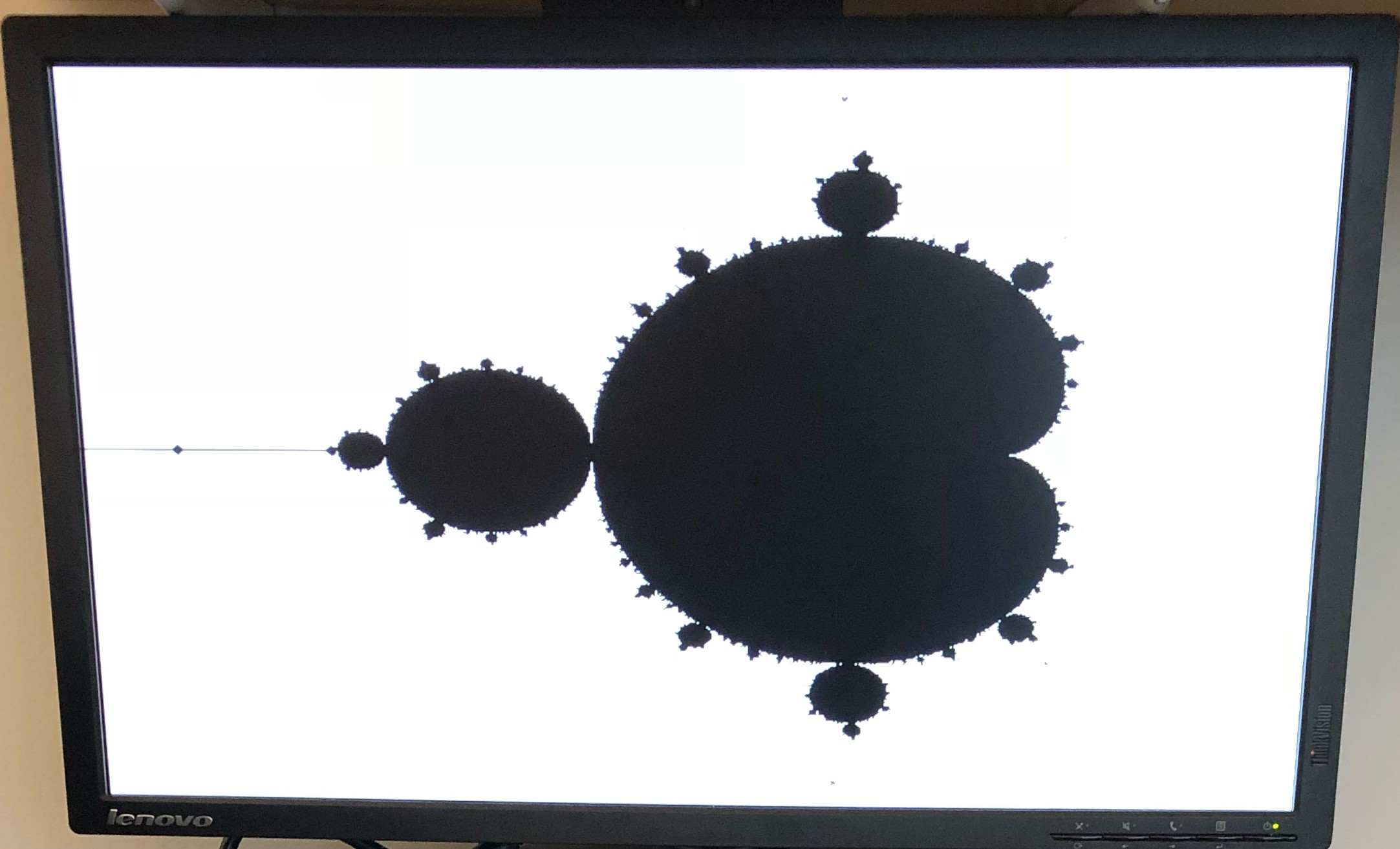} 
  }
    \vspace{-0.15in}
    \caption{Incorrectly (a) and correctly (b) rendered Mandelbrot set.}
    \label{mandelbrot}
    \vspace{-0.2in}
\end{figure}

The different versions of the processing cores in this platform lay the groundwork
for a quick micro-architecture design exploration. The cores included in
this platform provide various levels of complexity for users to
build on. In order to evaluate the ease of use of the platform, we instantiate a simple faulty core and add peripheral
devices such as a memory mapped VGA frame buffer. The frame buffer is used to
render a Mandelbrot fractal.
Testing the rendering program revealed that the fault was injected in the ALU module and caused it to execute 
signed operations incorrectly. Figure~\ref{mandelbrot} compares the incorrect
rendering with the correct rendering.
Finding and fixing this bug is an example of how this design space exploration
platform enables users to make additions easily and expand core designs.

\subsection{Design Space Exploration}

To evaluate the performance of the platform and showcase the extent of design space exploration
possible, we benchmark several processing system configurations.
First, we compare single cycle, five-stage pipeline (with and without data forwarding)
and seven-stage pipeline (with data forwarding) cores with asynchronous memory
in the single cycle system and synchronous memory in the pipelined systems.
Each system uses a dedicated instruction and data memory module in a single
core configuration. These configurations do
not use caches to avoid paying the penalty of cache misses without the benefit
of larger off-chip main memory.
Second, we compare multi-core systems connected via a bus between the level 1
and level 2 caches. In each multi-core system, the seven-stage pipelined core
is used.

To compare the single-core, cacheless processor configurations,
three different benchmark programs are run on each core. One benchmark computes
the factorial of an integer. The second benchmark counts the number of prime
numbers between two numbers. A third benchmark computes the
Mandelbrot set at a given resolution and checksums the result.

\begin{figure}[t]
    \centering
    \includegraphics[width=0.95\columnwidth]{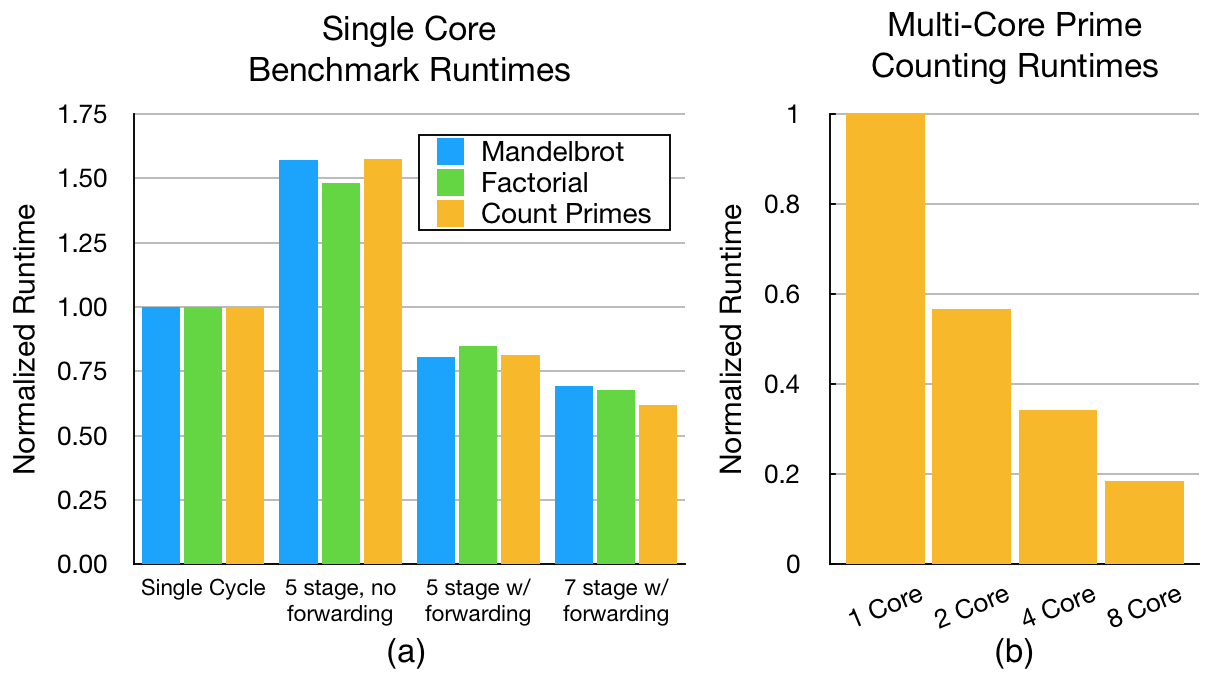}
    \vspace{-0.1in}
    \caption{(a) Runtime of each single-core system for each benchmark, normalized
      to the single cycle core runtime. (b) Runtime of single, dual, quad and
      octa-core processors for the prime number counting benchmark, normalized to
      the single core runtime.}
    \label{normal_runtime}
    \vspace{-0.1in}
\end{figure}

Each benchmark is executed in an RTL simulation of the configured processing
system. The number of cycles needed to complete the program execution is
recorded. The estimated $F_{\max}$ (obtained from synthesis tools)
of each core and the number of clock cycles in
each program execution is used to compute the runtime of the benchmark.
Table~\ref{runtime_table} reports the number of cycles for each benchmark.
Figure~\ref{normal_runtime} compares the runtime of each benchmark
on each of the configured systems. The runtimes are normalized to the single cycle
core's runtime.

The single cycle system executes each benchmark in the fewest cycles but the
low clock frequency hurts the program runtime. The single cycle core's clock
frequency is roughly half that of the other cores because the asynchronous
main memory is implemented in logic elements instead of the faster BRAMs.

\begin{center}
  \begin{table}[b!]
    \begin{tabular} {  | P{1.45cm} p{0.9cm} p{1.0cm} l l |  }
  \hline
  Core      & Clock   & Prime & Factorial & Mandelbrot \\
  Type      & Freq.  & Cycles    & Cycles    & Cycles     \\
  \hline
  Single Cycle                   & 29.0 MHz & 3,464k  & 59k  & 488k \\
  \hline
  5 Stage Stalled & 62.6 MHz & 11,789k & 190k & 1,654k \\
  \hline
  5 Stage Bypassed & 61.5 MHz & 5,891k & 107k & 833k \\
  \hline
  7 Stage Bypassed & 81.1 MHz & 6,833k & 113k & 948k \\
  \hline
  \end{tabular}
  \caption{Clock frequency and number of clock cycles to run each benchmark for
  each configured single core system.}
  \label{runtime_table}
  \vspace{-0.1in}
  \end{table}
\end{center}

\begin{center}
\begin{table}[b!]
  \centering
  \begin{tabular} { | P{2cm} P{3cm} P{2.1cm} |  }
  \hline
  Core  & Clock    & Number of \\
  Count & Frequency& Cycles    \\
  \hline
  1     & 61.7 MHz & 7,995,845 \\
  \hline
  2     & 59.5 MHz & 4,373,431 \\
  \hline
  4     & 60.0 MHz & 2,671,258 \\
  \hline
  8     & 59.5 MHz & 1,426,346 \\
  \hline
  \end{tabular}
  \caption{Clock frequency and number of clock cycles to run the prime counting
    benchmark for different number of cores.}
  \label{multicore_runtime_table}
\end{table}
\end{center}


The five-stage pipelined processor has a much higher clock frequency than the
single cycle core, but must
stall the pipeline for each hazard encountered. Pipeline stalls lead
to a much higher number of cycles needed to execute the program. The five-stage pipeline without data forwarding has the highest program runtime
of the tested cores.
Adding data forwarding to the five-stage pipeline cuts the program runtime in
half, yielding runtimes better than the single cycle core.

The seven-stage pipeline adds two stages to the five-stage pipeline to
support synchronous memories without inserting NOPs. These extra stages increase
the clock frequency but also increase the number of cycles needed to compute the
target address of branch and jump instructions.
The five-stage pipeline must wait two cycles before a jump or branch address
is ready while the seven-stage pipeline must wait three cycles.
Neither pipeline has a branch predictor. The extra cycles spent stalling
are canceled out by the higher clock frequency of the seven-stage pipeline.
The seven-stage pipeline has the best runtime for each benchmark.

\begin{figure}[t]
    \centering
    \includegraphics[width=0.80\columnwidth]{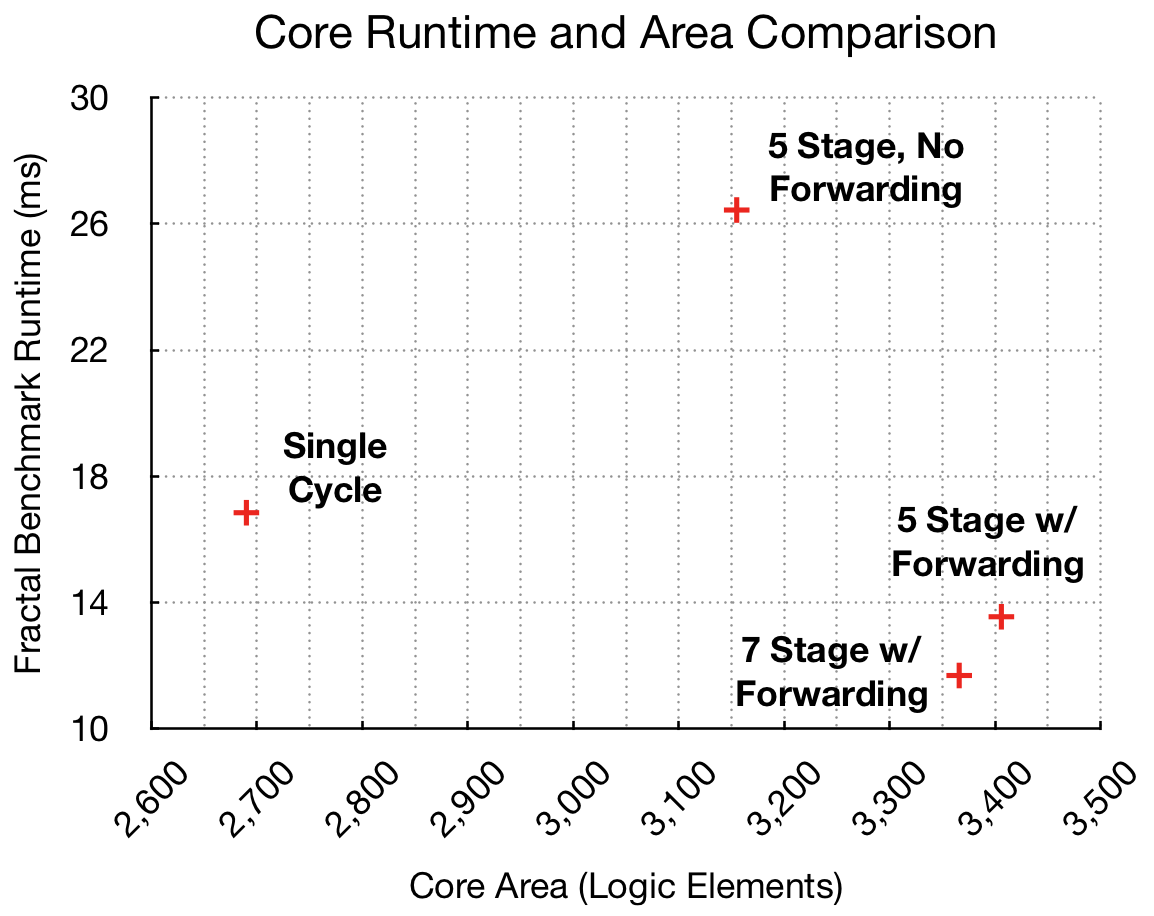}
    \vspace{-0.1in}
    \caption{Comparisons of the performance and area trade-offs made by each
    configured single-core system. Only core area is considered. Area used by memory is ignored.}
    \label{runtime_per_area}
    \vspace{-0.10in}
\end{figure}

Comparing the five-stage pipeline with and without
data forwarding demonstrates the effectiveness of forwarding in resolving
pipeline hazards. Comparing the five- and seven-stage pipelines with data
forwarding illustrates how the number of bubbles inserted in the pipeline during
jumps, branches, and the remaining load-use hazards impacts program runtime.
Figure~\ref{runtime_per_area} plots the Mandelbrot fractal benchmark runtime versus
the core area to visualize the area and performance trade-off. Only
the area of the core is considered; area used by memory is ignored here. The
area usage of each core is discussed in Section~\ref{sssec:synthesis_results}.

To compare multi-core architectures using caches, the same prime counting
benchmark used in the single core tests is parallelized and run on multi-core systems
with different core counts. The L1 instruction and data caches in the configured
multi-core systems have four 32-bit words per cache line and are 4-way set
associative. There are 256 lines in each L1 cache.
The shared L2 cache in each processor also has four 32-bit words per cache
line and is 4-way set associative. The L2 cache has 512 cache lines. As expected,
increasing the number of cores decreases the program runtime.
Table~\ref{multicore_runtime_table} shows the clock frequency of each multi-core
processor and the number of cycles needed to execute the benchmark program.
Figure~\ref{normal_runtime} compares the runtime of each processor.
Note that each time the number of cores is doubled, the runtime is nearly halved.

By providing different core options with a range of pipeline depths and clock
frequencies, the platform simplifies analysis of performance and area tradeoffs.
The analysis presented here was limited to core micro-architecture, cache design
and number of cores but many more options are available for fine tuning.

\subsection{Synthesis Results}\label{sssec:synthesis_results}
The runtime lengths presented in Tables~\ref{runtime_table}
and~\ref{multicore_runtime_table} are based on simulations, but each core is
fully synthesizable. All synthesis results presented here target an Altera
Cyclone IV FPGA with 150k logic elements.

Table~\ref{synthesis_results} shows the ``ADDRESS\_BITS'' parameter value, logic
element usage, BRAM usage and worst case $F_{\max}$ for each cacheless core type.
In Table~\ref{synthesis_results}, the ``ADDRESS\_BITS'' parameter was set to 12 for
each pipelined core. Twelve address bits yields 4k 32-bit word addresses for
each instruction and data memory (8k words total). The single cycle core
uses the asynchronous memory and must implement its memory in logic elements. To
ensure the single cycle design fits on the device, it was synthesized with just
1k word memories.

The cores in Table~\ref{synthesis_results} do not use caches. Separate instruction and data
RAMs are connected directly to the processors memory interfaces to act as main memory. Using
simple memory minimizes the memory system's impact on resource usage.
These results focus on the cores themselves. Tables~\ref{synthesis_results_cache_size},~\ref{synthesis_results_cache_associativity} and~\ref{synthesis_results_cache_linewidth}
show cache resource usage in isolation. Table~\ref{multicore_synthesis_results}
shows synthesis results for several multi-core processor configurations.

Table~\ref{synthesis_results_cache_size} shows synthesis results for varying
L1 cache sizes targeting the same Cyclone IV FPGA mentioned above. The number of
logic elements and BRAM used as well as the maximum clock frequency are
reported. Each cache is 4-way set associative and has 4 word (16 Byte) cache
lines. The synthesis results show that varying the cache size while maintaining
the same line size and associativity only marginally changes the logic element
usage and $F_{\max}$. Only the BRAM usage is significantly impacted by cache
size.

Table~\ref{synthesis_results_cache_associativity} shows synthesis results for
4kB L1 caches with 4 word (16 Byte), cache lines and various associativities.
Intuitively, the number of logic elements grows with the associativity while
$F_{\max}$ shrinks. The BRAM usage grows slowly with the associativity. The
16-way associative cache uses 2k more BRAM bits than the 1-way
(direct mapped) cache. The increase in BRAM usage with higher associativities
can be explained by the higher BRAM usage in the `replacement\_controller' module,
which implements the true least recently used (LRU) replacement policy.

Table~\ref{synthesis_results_cache_linewidth} shows synthesis results for 4kB
L1 caches with line widths ranging from 1 word (4 Bytes) to 16 words (64 Bytes).
Varying the line width trades off BRAM and logic elements. Smaller line widths
use more BRAM bits for tag storage, while larger line widths use more logic
elements to implement registers and buses to handle wider cache lines.

Table~\ref{multicore_synthesis_results} shows the resource usage for processors
with 1, 2, 4 and 8 cores. The cores used in the multi-core processor are
versions of the 7-stage pipelined core. The L1 caches, shared L2 cache, and coherence
controller modules add a significant area overhead but enable the use of complex
memory hierarchies. The significantly larger BRAM usage in the multi-core
processors stems from the large cache and on-chip main memory. The memory in
the single core processors was made to be smaller because larger memories were
not needed for the benchmarks. Note that the BRAM usage for the 8-core processor
is slightly greater than what is available on the device. We have included the
results here because moving the main memory to off-chip SRAM available on our
development board would allow the caches to fit within the device BRAM without
significantly impacting the performance.

The memory hierarchy for each of the processors in Table~\ref{multicore_synthesis_results}
uses the same parameters. 16kB L1 caches, 32kB L2 caches and a main memory of 256kB.

The different cores and cache configurations supported by the platform enable users to
examine the difference between design choices quickly. Additional configuration
options, including NoC routing, NoC topology, and on/off-chip main memory configurations
have not been included here due to space restrictions.

Each core builds off of the same set of base modules, making it easy to integrate
experimental features into several of them by modifying only the base module.
This extensibility opens up even more possibilities for design space exploration.
We have used this technique to add data forwarding, originally
added to the five-stage pipeline, to the out-of-order and seven-stage pipeline
cores.

\begin{center}
  \begin{table}[t]
    \begin{tabular} {  |P{1.5cm} p{0.6cm} p{1.7cm} p{1.50cm} p{1.10cm}|  }
  \hline
  Core      & Addr & Logic & BRAM & Fmax \\
  Type      & Bits & Elements & Bits & \\
  \hline
  Single Cycle & 10   & 53,448   & 0            & 29.0 MHz\\
  \hline
  5 Stage Stalled & 12 & 3,160 & 262,144 & 62.6 MHz \\
  \hline
  5 Stage Bypassed & 12 & 3,406 & 262,144 & 61.5 MHz \\
  \hline
  7 Stage Bypassed & 12 & 3,366 & 262,144 & 81.1 MHz \\
  \hline

  \end{tabular}
  \caption{Synthesis results for each configured single-core processing system.}
  \label{synthesis_results}
  \vspace{-0.10in}
  \end{table}
\end{center}

\begin{table}[t]
  \centering
  \begin{tabular} {  |P{1.4cm} p{1.6cm} p{2.35cm} p{1.5cm}|  }
    \hline
    Core  & Logic    & Total Memory & Fmax      \\
    Count & Elements & Bits (BRAM)  & Fmax      \\
    \hline
    1     & 12,206   & 2,736,128 & 61.7 MHz\\
    \hline
    2     & 21,123   & 3,055,616 & 59.5 MHz \\
    \hline
    4     & 38,921   & 3,694,592 & 60.0 MHz \\
    \hline
    8     & 83,020   & 4,972,544 & 59.5 MHz \\
    \hline
  \end{tabular}
\caption{Synthesis results for each configured multi-core processing system.}
\label{multicore_synthesis_results}
\vspace{-0.10in}
\end{table}

\begin{center}
  \begin{table}[t!]
    \centering
    \begin{tabular} {  |p{0.9cm} p{0.7cm} p{1.3cm} p{2.2cm} p{1.5cm}|  }
  \hline
  Cache & Addr & Logic    & Total Memory & Fmax       \\
  Size  & Bits & Elements & Bits (BRAM)  & \\
  \hline
      1kB & 4 & 3,208 & 10,240 & 79.1 MHz \\
  \hline
      2kB & 5 & 3,196 & 20,352 & 81.0 MHz \\
  \hline
      4kB & 6 & 3,198 & 40,448 & 82.6 MHz \\
  \hline
      8kB & 7 & 3,189 & 80,384 & 81.1 MHz \\
  \hline
      16kB & 8 & 3,189 & 159,744 & 82.8 MHz \\
  \hline
  \end{tabular}
  \caption{Synthesis results for various sizes of 4-way set associative L1 caches.
    Each cacheline is 16 Bytes (four 32-bit words)}
  \label{synthesis_results_cache_size}
  \vspace{-0.05in}
  \end{table}
\end{center}

\begin{center}
  \begin{table}[t!]
    \begin{tabular} {  |p{1.1cm} p{0.7cm} p{1.1cm} p{2.2cm} p{1.5cm}|  }
  \hline
  Cache & Addr & Logic    & Total Memory & Fmax       \\
  Ways  & Bits & Elements & Bits (BRAM)  & \\
  \hline
      1 & 8 & 2,387 & 39,424 & 104 MHz \\
  \hline
      2 & 7 & 2,736 & 39,936 & 97 MHz \\
  \hline
      4 & 6 & 3,198 & 40,448 & 82 MHz \\
  \hline
      8 & 5 & 4,534 & 40,960 & 66 MHz \\
  \hline
      16 & 4 & 6,546 & 41,472 & 49 MHz \\
  \hline
  \end{tabular}
  \caption{Synthesis results for various associativities in 4kB L1 caches.
    Each cacheline is 16 Bytes (four 32-bit words)}
  \label{synthesis_results_cache_associativity}
  \vspace{-0.10in}
  \end{table}
\end{center}

\begin{center}
  \begin{table}[t!]
    \begin{tabular} {  |p{1.4cm} p{0.7cm} p{1.1cm} p{2.2cm} p{1.3cm}|  }
  \hline
  Line    & Addr & Logic    & Total Memory & Fmax\\
  width         & Bits    & Elements & Bits (BRAM)  & \\
  \hline
      4 Bytes  & 8 & 1,663 & 63,488 & 82 MHz \\
  \hline
      8 Bytes  & 7 & 2,227 & 48,128 & 81 MHz \\
  \hline
      16 Bytes & 6 & 3,198 & 40,448 & 82 MHz \\
  \hline
      32 Bytes & 5 & 5,039 & 36,608 & 81 MHz \\
  \hline
      64 Bytes & 4 & 8,888 & 34,688 & 76 MHz \\
  \hline
  \end{tabular}
  \caption{Synthesis results for various line widths in 4kB, 4-way set associative L1 caches.}
  \label{synthesis_results_cache_linewidth}
  \end{table}
\end{center}

\vspace{-0.4in}
\section{Future Work}
The goal of this work is to explore techniques to develop a fast, flexible,
multi-core design space exploration platform, enabling users to understand the
impact of their design decisions and quickly test different configurations.
To expand the number of configuration options available to users further, the
authors plan to add several new features to the available cores, including branch predictors, hardware multi-threading, and other RISC-V ISA extensions,
such as the floating point (RV32F) or multiply (RV32M) extensions. Each of these
new features will be made available as an option for a user's desired core and
incorporated into the hardware configuration GUI.

The platform has already supported, and will continue to support, research
relating to secure architectures focused on enabling efficient obfuscation with hardware-software
co-design. The improvements described above will be necessary to support
additional research focused on developing RISC-V architectures for HPC workloads
with efficient distributed memory.

As a design space exploration platform, the explorable design space is only
limited by the man-power available to develop configurable features. For this
reason, we have released the source code for the entire platform (the RTL,
toolchain customizations and GUI application code) in the hope that others in
the community can benefit from, and add to the design space exploration capabilities of the tool.

\section{Conclusion}
This platform works to address the challenge of fast multi-core design space
exploration.
By offering highly parameterized cores, cache, memory, and NoC subsystems,
our platform allows users to quickly
explore a RISC-V architectural design space without the need to develop complex
multi-core systems from scratch.
A supporting hardware configuration application GUI enables rapid selection of
system parameters and RTL generation. Once Verilog is generated by the
configuration application, users can investigate how each design decision will impact system
properties such as performance, area or timing. Users can add custom features or other
modifications to further expand the explorable design space.

Including the necessary
compiler tool-chain makes running experiments on customized hardware systems
simpler. The whole platform system is open source, including all of the RTL code, \\
toolchain customizations and supporting applications, enabling \\users to customize
components to fit their needs.

\bibliographystyle{ACM-Reference-Format}
\bibliography{paper} 

\end{document}